\documentclass[epj]{svjour}
\usepackage{graphics}
\begin{document}
\title{Wetting of cholesteric liquid crystals}
\author{Nuno M. Silvestre\inst{1} \and Maria Carolina Figueirinhas Pereira\inst{1} \and Nelson R. Bernardino\inst{1} \and Margarida M. Telo da Gama\inst{1}
}                     
\institute{Centro de F\'{\i}sica Te\'{o}rica e Computacional and Departamento de F\'{\i}sica, Faculdade de Ci\^{e}ncias, Universidade de Lisboa, 1749-016
Lisboa, Portugal.}
\date{Received: date / Revised version: date}
%
\abstract{
We investigate theoretically the wetting properties of cholesteric liquid crystals at a planar substrate. If the properties of substrate and of the interface are such that the cholesteric layers are not distorted the wetting properties are similar to those of a nematic liquid crystal. If, on the other hand, the anchoring conditions force the distortion of the liquid crystal layers the wetting properties are altered, the free cholesteric-isotropic interface is non-planar and there is a layer of topological defects close to the substrate. These deformations can either promote or hinder the wetting of the substrate by a cholesteric, depending on the properties of the cholesteric liquid crystal. %
%
} 
\maketitle
\section{Introduction}
\label{intro}
The understanding of surface and interfacial properties of fluids dates back to the early 19th century work of Young \cite{Rowlinson1982} but continues to play a prominent role in modern science. When the fluid is a liquid crystal the phenomenology of interfacial phenomena and wetting is very rich and the unique features of liquid crystals, such as elasticity and the presence of topological defects, are promising routes to new microfluidic devices, where interfacial properties play a critical role \cite{Sengupta2011}.

Alongside the technological applications the understanding of the wetting properties of liquid crystals \cite{deGennes1993} are also a key fundamental problem. The study of the wetting behavior of the simplest of liquid crystals, nematics, at planar substrates has a long history (for a review see \cite{Jerome1991}). 
It is known for a long time that the wetting properties of nematic liquid crystals have similarities with those of simple fluids, as long as the surfaces do not impose elastic distortions on the liquid crystal \cite{Sheng1976,Sheng1982,Yokoyama1983,Hsiung1986,Sen1987,Hsiung1987,Chen1989,Alkhairalla1999}. If elastic distortions are present the wetting properties are dominated by elasticity and are thus very different but well understood \cite{Braun1996} .

The unique features of liquid crystals are more prominent when the substrate is chemically or geometrically patterned. In this case the presence of topological defects has a profound influence on the wetting properties \cite{Bramble2007,Harnau2004,Patricio2008,Romero-Enrique2010,Patricio2011,Patricio2011a,Rojas-Gomez2012,Silvestre2012}, such as re-entrant wetting, the suppression of wetting or filling, etc.

In contrast to nematics, very little is known about the theory of wetting by cholesterics, and even experimental results have been published only recently \cite{Zola2013}. Cholesteric liquid crystals, also known as chiral nematics, are nematic liquid crystals in which the preferred alignment direction describes a helix with a given pitch along an axis. In many ways cholesteric liquid crystals behave as layered materials \cite{deGennes1993}. The absence of theoretical studies of wetting of cholesterics is not a surprise since even the cholesteric-isotropic interface was understood only recently\cite{Bernardino2014}. It was  well known for some time that the cholesteric-isotropic interface might not be planar, exhibiting instead a series of parallel ridges and valleys \cite{Meister1996a,Meister1996,Saupe1973,Yoshioka2014,Zola2013,Rofouie2014,Rofouie2015}. This rough interface rules out the possibility of using one-dimensional calculations that are usual in theoretical studies of wetting and makes wetting of a cholesteric at a planar substrate resemble somewhat wetting of a nematic at a patterned 
substrate, requiring a more sophisticated numerical analysis. 
Our recent systematic study of the free cholesteric-isotropic interface confirmed the non-planar configuration of the interface and revealed a non-trivial dependence of the surface tension on the pitch and the elastic 
properties of the liquid crystal \cite{Bernardino2014}. The goal of this manuscript is to contribute to fill this gap in our knowledge of the fundamental wetting properties of cholesteric liquid crystals.

In the next section we will detail our model and the numerical methods. In section 3 we present our results, starting with a brief review of wetting phenomena, of the properties of the cholesteric isotropic interface, and some analytical results. The remainder of section 3 describe how the wetting properties of the cholesteric depend on the pitch and on the elastic properties of the liquid crystal. Finally in the last section we summarise our main findings and hint at some open questions.

\section{Model}
\label{sec_model}

Our analysis follows a well established route, using the Landau-de Gennes free energy for a cholesteric liquid crystal (LC) \cite{deGennes1993,Patricio2011}, which is built from the usual expansion in the lowest order terms of the symmetric, traceless, tensor order parameter $Q_{ij}$, split into elastic $f_{e}$, bulk $f_{b}$, and surface terms $f_{s}$, $F = \int d V \left( f_{e} + f_{b}\right) + \int d S f_{s}$. Here $f_{e} = \frac{1}{3+2 k} ( Q_{ij,k} Q_{ji,k} +  4q_0 Q_{il}\epsilon_{ijk} Q_{kl,j} + 4q_0^2 Q_{ij}Q_{ji} +k Q_{ij,j}Q_{ki,k})$, $f_{b} = \frac{2}{3}\tau Q_{ij}Q_{ji} - \frac{8}{3}Q_{ij}Q_{jk}Q_{ki} + \frac{4}{9}(Q_{ij}Q_{ji})^2$, and $f_{s} = -\frac{2}{3}\omega Q_{ij}Q_{ji}^s$ where summation over repeated indices is assumed. Our version uses dimensionless quantities and thus the bulk free-energy depends only on three parameters, $q_0$, $\tau$, and $k$.
$q_0\equiv{2\pi}/{P}$ is the inverse of the cholesteric pitch $P$. In this model the nematic phase is described by the limit of infinite pitch $P\to\infty$ or, equivalently, the limit $q_0\to 0$.
 $\tau$ is a reduced temperature whose value determines the equilibrium bulk phase.
While for a nematic $\tau=1$ is the coexistence temperature between the nematic and the isotropic phases, for a cholesteric the coexistence temperature depends both on $k$ and the pitch: 
$ \tau_c = 1 - {3 q_0^2}/{(6+4k)}$.  
The order in the LC phase is described by a scalar order parameter $S$ with values $S=0$ in the isotropic phase and $S=S_b$ in the LC phase. At the coexistence temperature, $S_b\approx 1$.
 All lengths are measured in units of the correlation length $\xi$, which is the scale of the typical size of the topological defects and of the width of the LC-isotropic interface (typically a few correlation lengths). As a reference, for the nematic LC 5CB the correlation length at room temperature is around $15$ nm so $P=1000\xi$ is equivalent to $P=15 \mu$m.
$k\equiv{L_2}/{L_1}$, where $L_1$ and $L_2$ are the usual elastic constants of the Landau-de Gennes theory. The usual elastic constants of the Frank-Oseen theory $K_1$ (splay), $K_2$ (twist), and $K_3$ (bend) are related to $S_b$, $L_1$, and $L_2$ by $K_1=K_3=9S_b^2(2L_1+L_2)/{4}$ and $K_2=9S_b^2{L_1}/{2}$. Notice that this simplest version of the Landau-de Gennes theory only has two elastic constants and $K_1=K_3$. Also notice that $k=0$ is equivalent to $L_2=0$, and $K_1=K_2=K_3$, known as the one-elastic-constant approximation.

The surface term favours a configuration of the tensor order parameter of $Q_{ji}^s$ near the surface, with $\omega$ being known as the anchoring strength. A value of $\omega$ near zero models a substrate with weak anchoring and the opposite limit of large values of $\omega$ models a substrate with strong anchoring, imposing a fixed orientation.

For the numerical results we assume translational invariance along the $z$ direction and thus calculate the configuration on the $xy$ plane. The Landau-de Gennes free energy is minimized using a Finite Element Method with adaptive meshing to resolve the different length scales \cite{Patricio2002}.The obtained numerical precision in free energy was lower than $1\%$.

\section{Results}
\subsection{Review of wetting phenomena}

When a liquid and its vapor are in contact with a solid surface it forms a drop with a defined contact angle $\theta$, whose value is given by Young's law, relating the value of the contact angle with the values of the surface tensions of the three interfaces that meet at the contact line: solid-vapor $\sigma_{sv}$, solid-liquid $\sigma_{sl}$, and liquid-vapor $\sigma$
\begin{equation}
\cos\theta = \frac{\sigma_{sv}-\sigma_{sl}}{\sigma}
\end{equation} 
From Young's equation we can see that the contact angle goes to zero when:
\begin{equation}
\sigma_{sv} = \sigma_{sl}+\sigma
\end{equation}
This condition, known as Antonow's rule, is the macroscopic equivalent of the microscopic condition that a film of adsorbed liquid has an infinite thickness and it defines a wetting transition, a surface phase transition that separates regimes of disparate interfacial properties. The approach to the wetting transition is controlled by three thermodynamic parameters: temperature, pressure and a surface field that encapsulates the influence of inter-molecular forces and surface chemistry. From a theoretical point of view the study of wetting phenomena is most transparent by holding the pressure and temperature fixed and changing the surface field.

For a liquid crystal there are further thermodynamic variables but wetting is easier to understand by setting the temperature to the coexistence temperature between the liquid crystal and the isotropic phases and changing the surface properties. Note that pressure does not play a role in the Landau-de Gennes model. The surface properties can be further split into anchoring (alignment of the director at the surface) and a surface field that controls the strength of the surface interactions $\omega$, the anchoring strength. We have already introduce this parameter before when we defined the theoretical model.

Specializing the above framework to cholesteric liquid crystals, Antonow's rule for wetting of a substrate by a cholesteric liquid crystal is 
\begin{equation}
\sigma_{si}(\omega_c) = \sigma_{sc}(\omega_c) + \sigma,
\label{eq_WettingCondition}
\end{equation}
where $\sigma_{si}, \sigma_{sc},$ and $\sigma$ are the surface tensions of the substrate-isotropic, substrate-cholesteric, and isotropic-cholesteric interfaces . The transition anchoring constant $\omega_c$ is the value of $\omega$ that satisfies this condition and defines the thermodynamic location of the wetting transition.

\subsection{Review of the cholesteric-isotropic interface}

\begin{figure}

\resizebox{\columnwidth}{!}{%
  \includegraphics{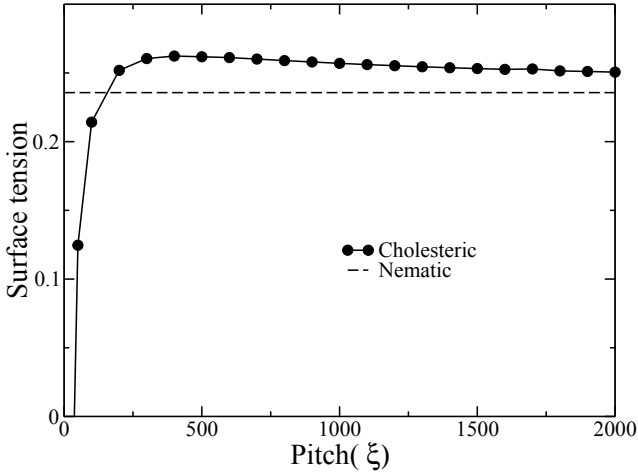}
}

\caption{Surface tension of the cholesteric-isotropic interface with the pitch, for $k=-1$. The surface tension decreases for lower values of the pitch, due to the formation of low free energy regions of double twist close to the interface. The value of the surface tension slowly approaches the nematic limit for large values of the pitch.}
\label{fig_surftension_pitch}       
\end{figure}

A key ingredient of wetting is the free cholesteric-isotropic interface. We described the structure and interfacial properties of the free cholesteric-isotropic interface recently \cite{Bernardino2014} but we review the main results here for completeness. Let us start with a nematic-isotropic interface. Using a simple ansatz \cite{deGennes1971,Patricio2008} it can be seen that the anchoring at the interface depends on the value of $k$: when $k>0$ the director aligns parallel to the interface (planar anchoring), whereas if $k<0$ the liquid crystal aligns perpendicular to the interface (homeotropic anchoring). For a cholesteric, planar anchoring is favoured if $k>k_c(P)$ where $k_c$ depends on the value of the pitch and $k_c\to 0$ as the pitch goes to infinity, the nematic limit \cite{Bernardino2014}. In this case the surface tension of the cholesteric-isotropic interface is the same as the nematic-isotropic interface. If $k<k_c(P)$ the liquid crystal tries to have 
homeotropic anchoring but this is not compatible with the layered structure of the cholesteric. To accommodate these incompatible 
configurations the cholesteric layers distort near the interface, forming topological defects, and the interface bends. The surface tension varies with the pitch, having a maximum and then decreasing slowly to the value of the surface tension of the nematic as the pitch increases, see Fig.~\ref{fig_surftension_pitch}. The deformations of the interface grow linearly with $|k|$ for $k<k_c(P)$ and grow sub-linearly with $\sqrt{P}$.

\begin{figure}
\resizebox{\columnwidth}{!}{%
  \includegraphics{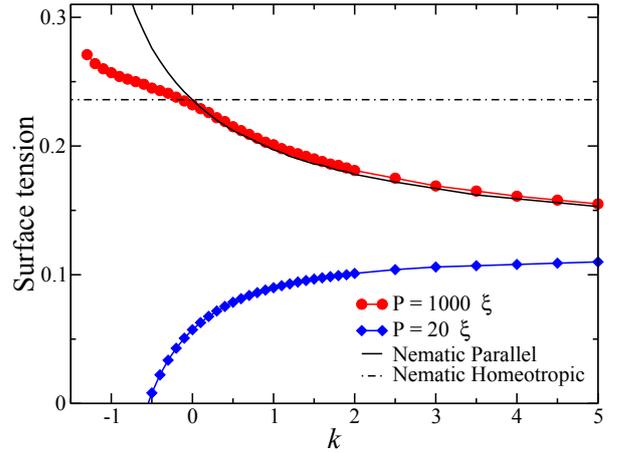}
}

\caption{Surface tension of the cholesteric-isotropic interface with $k$, for pitch=$20\xi$ (blue diamonds) and pitch=$1000\xi$ (red dots). The surface tension of the nematic-isotropic interface for parallel (stricht black line) and homeotropic (dashed black line) anchoring is also plotted for reference.}
\label{fig_surftension_kappa}       
\end{figure}

\subsection{Planar anchoring at the cholesteric-isotropic interface}

The simplest wetting scenario with a cholesteric liquid crystal is when $k>k_c$ and thus the anchoring at the interface is planar. If the anchoring at the substrate is also planar the layered structure of the cholesteric can be accommodated in a wetting layer without distortions. Since there are no distortions the problem of calculating the equilibrium configuration and the corresponding surface tension is effectively one-dimensional and we  can calculate the wetting anchoring constant using the same ansatz used for nematics. As we already know the surface tension is the same as for nematics and independent of the value of the pitch. This is consistent with the view that at length scales smaller than $P$ the cholesteric is similar to a nematic. The substrate-cholesteric and substrate-vapor surface tensions can be calculated with the same ansatz and using Antonow's rule we obtain the same value for $\omega_c$ as for the nematic:
\begin{equation}
\omega_c = \sigma = \frac{1}{6} \sqrt{\frac{6+k}{3+2k}}.
\label{eq_WettingCondition2}
\end{equation}

We checked these results numerically and they are consistent with the theoretical prediction, within $4 \% $. We note that the theoretical calculations assume that the configuration of the cholesteric is not changed when $S$ varies close to the interface and biaxiality is ignored. These assumptions might be responsible for the discrepancy between the theoretical and the numerical results. As an example if we allow for biaxiality we obtain for the coexistence temperature between isotropic and cholesteric $\tau_c = 1 - 3q_k^2 - \left(1 -\left(1 + 2q_k^2\right)^{3/2}\right)/2$, where $q_k \equiv {q_0}/{\sqrt{3+2k}}$. This is not significantly different from the value neglecting biaxiality $ \tau_c = 1 - 3 q_0^2/\left(6+4k\right)$ except for very small values of the pitch.

As a summary, in the ``non-distorted'' case, the cholesteric and the nematic have the same wetting properties. The only important distinction is that the nematic favours planar anchoring for $k>0$, whereas for the cholesteric this happens for $k>k_c(P)$, where $k_c(P)$ depends on the pitch and is positive.

\subsection{Homeotropic anchoring at the cholesteric-isotropic interface: dependence on the pitch}
\label{sec_results_2}

\begin{figure}
\resizebox{\columnwidth}{!}{%
  \includegraphics{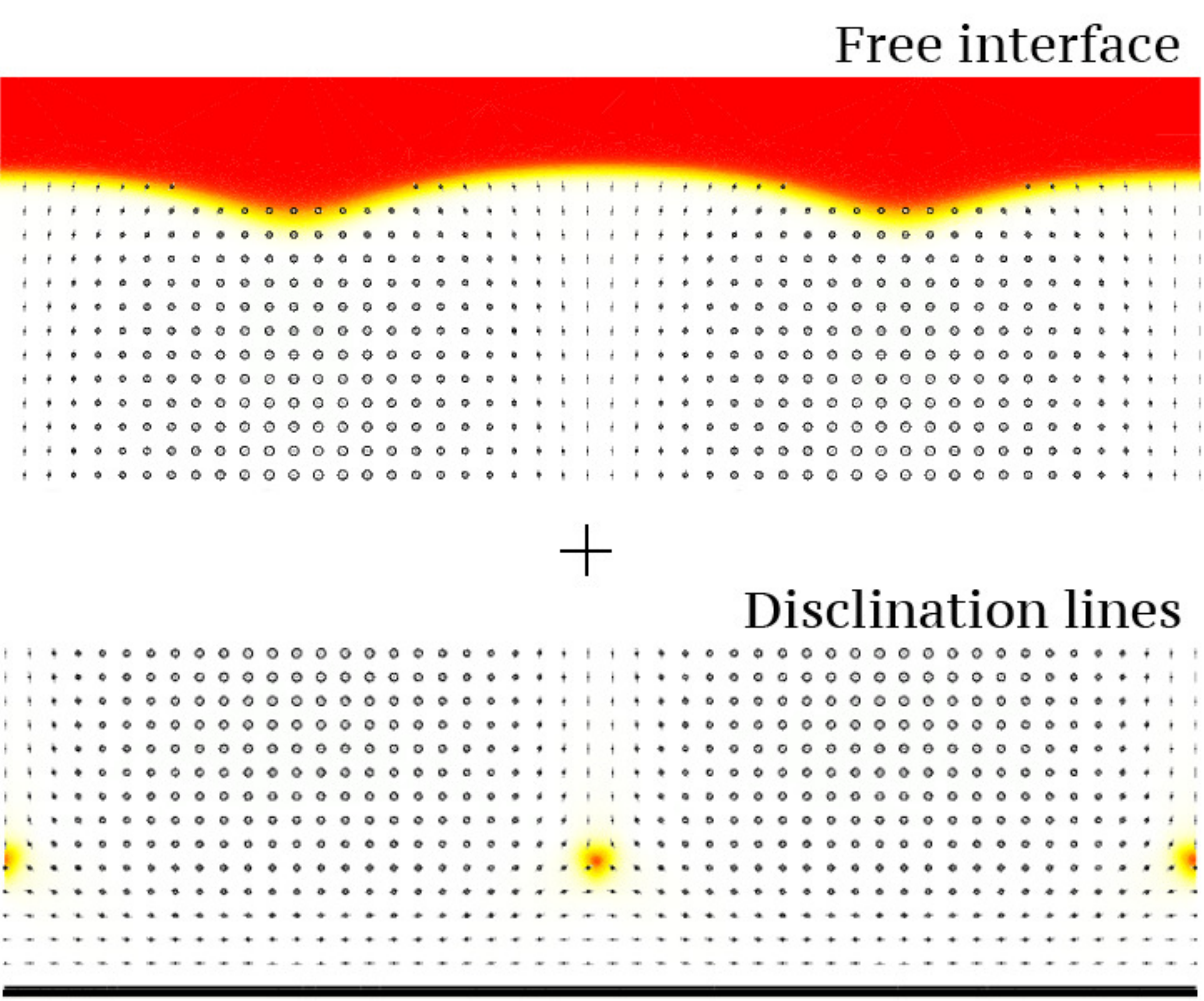}
}
\caption{Typical configuration of a wetting layer with negative $k$ showing the nonplanar interface and the layer of topological defects close to the substrate. The isotropic phase is plotted in red and the circles and lines give the direction of the director in the cholesteric phase. A large circle indicates a large out-of-plane director component.}
\label{fig_config}       
\end{figure}

Unlike the case $k>k_c(P)$ described above, for systems with $k<k_c(P)$ things are more interesting. The cholesteric layers prefer to align perpendicular to the interface and the planar anchoring at the substrate may force the nucleation of a further layer of topological defects close to the substrate, so that all topological constraints are satisfied. A typical configuration is seen in Fig.~\ref{fig_config}.

\begin{figure}
\resizebox{\columnwidth}{!}{%
  \includegraphics{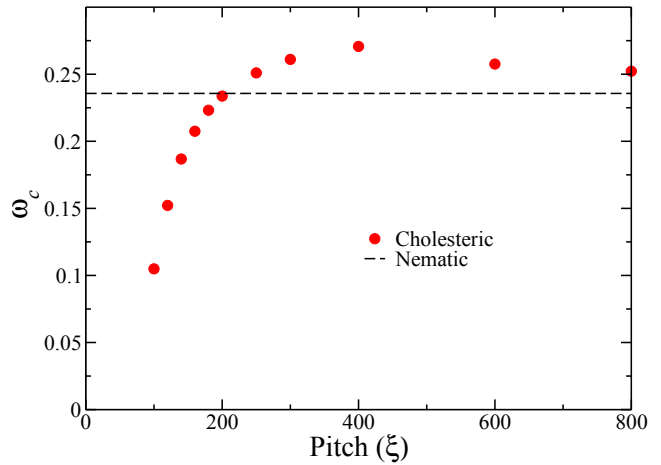}
}
\caption{Critical anchoring $\omega_c$ as a function of the pitch $P$, for $k=-0.5$. As a reference the value for a nematic is presented.}
\label{fig_pitch}       
\end{figure}

In Fig.~\ref{fig_pitch}, we plot the value of the critical anchoring $\omega_c$  as a function of $P$ for a negative value of $k = -0.5$. For comparison, we also plot $\omega^c$ for the nematic. As expected, the overall behavior of $\omega_c$ follows the surface tension for $k <0$ shown in Fig.~\ref{fig_surftension_kappa}. For small $P\leq 200\xi$ $\omega_c$ decreases as $P\to 0$. The underlying cause of this behavior is the same as for the surface tension: the formation of double-twist regions ($\lambda$ defects) close to the cholesteric-isotropic interface with lower free energy density than the bulk cholesteric phase, favouring wetting of the substrate by the cholesteric phase. On the opposite limit, for large values of $P$, $\omega_c$ approaches slowly the asymptotic value (the critical anchoring for a nematic). The difference between the values of the critical anchoring of the cholesteric and of the nematic is due to the undulations of 
the isotropic-cholesteric interface, the distortions of the cholesteric layers and the formation of  topological defects near the substrate. This layer of topological defects has a strong influence on the wetting behavior which is not obvious from the previous result. In the next section we will see how $\omega_c$ depends on $k$, which highlights the role of the topological defects.

\subsection{Homeotropic anchoring at the cholesteric-isotropic interface: dependence on $k$}

\begin{figure}
\resizebox{\columnwidth}{!}{%
  \includegraphics{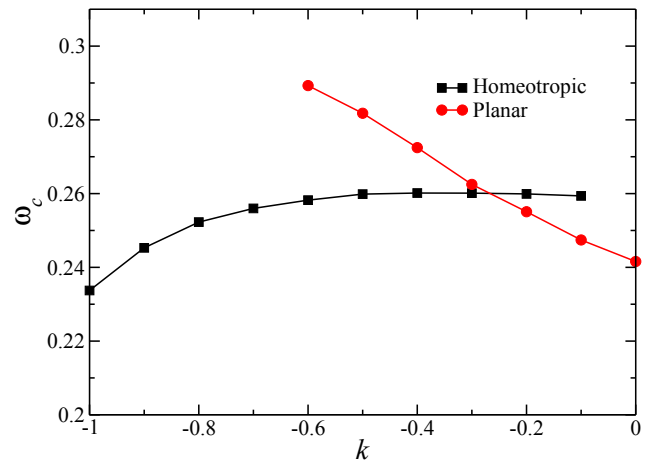}
}
\caption{Critical anchoring $\omega_c$ as a function of $k$, for pitch  $P=200\xi$. The preferred wetting state changes from homeotropic to planar at $k\simeq -0.3$.}
\label{fig_k_small}       
\end{figure}

\begin{figure}
\resizebox{\columnwidth}{!}{%
  \includegraphics{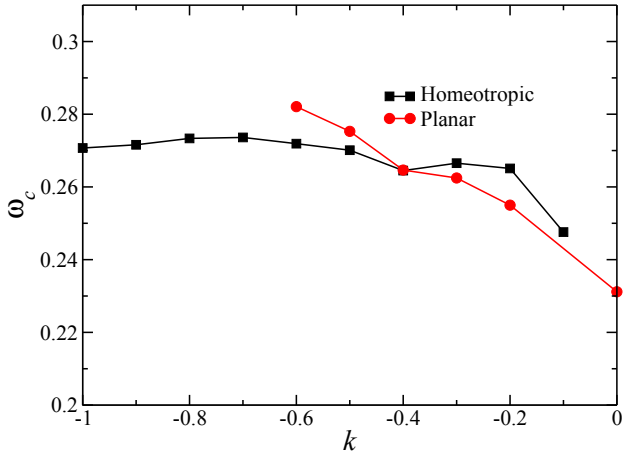}
}
\caption{Critical anchoring $\omega_c$ as a function of $k$, for pitch $P=400\xi$. The preferred wetting state changes from homeotropic to planar at $k\simeq -0.4$.}
\label{fig_k_big}       
\end{figure}

In this section we explore the role of $k$ on the behavior of $\omega_c$ taking into account the value of the pitch. In Fig.~\ref{fig_k_small} and Fig.~\ref{fig_k_big} we plot $\omega_c$ with $k$ for cholesteric layers aligned parallel and perpendicular to the interface for $P=200\xi$ and $P=400\xi$, respectively.

In both Fig.~\ref{fig_k_small} and Fig.~\ref{fig_k_big}, the curves of $\omega_c$ for the cholesteric cross at a negative value of $k$ ($k_c \approx -0.3$ for $P=200\xi$ and $k_c \approx -0.4$ for $P=400\xi$), whereas the curves for the nematic cross at $k =0$, i.e. the presence of the substrate causes a change of the value of $k_c(P)$ at which the preferred anchoring changes from planar to homeotropic. This change is due to the presence of the layer of topological defects close to the substrate, which penalizes homeotropic anchoring. As the value of the pitch increases the value of the free-energy per pitch of the layer of topological defects decreases to zero and $k_c(P) \to 0$, the nematic limit. Note that $k_c(P)$ for the free isotropic-cholesteric interface is also different from zero \cite{Bernardino2014} due to the double-twist regions close to the interface.

\begin{figure}
\resizebox{\columnwidth}{!}{%
  \includegraphics{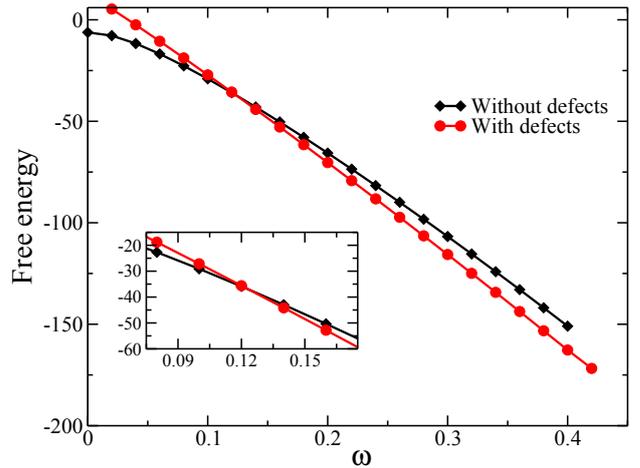}
}
\caption{Free energy of the configurations with (red) and without (black) topological defects near the surface, as a function of the anchoring strength $\omega$. The crossing identifies when the layer with defects is formed. The inset shows the transition, which takes place at $\omega\simeq0.12$.}
\label{fig_prewet}       
\end{figure}

In the results described above we have considered that the wetting configuration had a layer of topological defects close to the substrate. This is what happens in the limit of strong anchoring but for smaller values of the anchoring there is a second possible configuration of the cholesteric without topological defects close to the substrate. In other words, there are two possible wet states. It may happen that for different parameters the formation of topological defects followed by a wetting transition is reversed. If this happens the wet state will form initially without topological defects and only for larger values of the anchoring will the topological defects be present. In order to see if the wet state at the wetting transition contains the disclination lines close to the substrate, we calculated numerically the value of the anchoring strength at the substrate for which the free energies of the systems with and without the additional defects are equal $\omega^*$ (see Fig.~\ref{fig_prewet}). 

We started with a system filled with a cholesteric in contact with a substrate that imposes parallel anchoring (at the bottom of the system). The cholesteric layers were set to be aligned perpendicularly to the substrate by imposing fixed anchoring at the top of the system. We increased the value of $\omega$ at the substrate until the disclination lines formed. The free energy of the configurations without defects for the different values of $\omega$ are plotted as the red curve in Fig.~\ref{fig_prewet}. After the formation of the defects, we decreased the value of $\omega$, obtaining the blue curve in Fig.~\ref{fig_prewet}. The crossing point of the red and blue curves is the value of $\omega^*$. Comparing $\omega^*$ with $\omega_c$ we see that $\omega^* < \omega_c$, i.e., the formation of the disclination lines occurs before the wetting transition. Hence, the wet state at the wetting transition contains the additional disclination lines. This is the most common scenario but it is possible that 
the reverse happens for very small values of the pitch or very negative values of $k$.

\section{Conclusions and discussion}

In this manuscript we studied wetting with a fixed orientation imposed at the substrate: planar and along the $x$ axis. We have also explored what happens for other values of the orientation at the substrate, when the preferred orientation is still planar but along the $z$ axis or at an intermediate direction between the $x$ and $z$ axis. We found that, for most values of the parameters, the state with the lowest energy is the one with alignment along the $x$ axis, even though for small values of the pitch or $k$ alignment along the $z$ axis may be favoured.

It is also possible that a substrate imposes degenerate planar anchoring, where the director is in the plane of the substrate but no direction in the plane is favoured, homeotropic anchoring, or oblique anchoring. Our goal in this manuscript is to describe the basic wetting phenomena of cholesterics rather than an exhaustive description. With the large number of parameters an exhaustive mapping of cholesteric wetting is a task for further work.

As a summary, we have described the main characteristics of wetting of a substrate by a cholesteric. The number of parameters makes an exhaustive exploration of the parameter space a gigantic task but we were able to extract the main features. Unlike a nematic liquid crystal wetting of a cholesteric at a planar substrate presents a rich phase diagram. The wetting characteristics follow broadly the same pattern, being dominated by the behaviour of the surface tension of the cholesteric-isotropic interface and can be divided into two regimes, for small and large values of the pitch. The topological constraints may force the nucleation of a surface layer of topological defects that we predict is nearly always present in the wet state.

It will be interesting to explore wetting phenomena of cholesterics at patterned substrates. Wetting at patterned substrates has been something of a hot topic in 
surface science (\cite{Patricio2011} and references therein) but cholesterics introduce a new feature in the phenomena. Since cholesterics have an intrinsic mesoscopic length scale, the pitch, we expect a rich interplay between the wavelength of the patterns at the substrate and the pitch of the cholesteric. It is not hard to anticipate that the ratio of these length scales will lead to phenomena such as frustration, elastic strains and a kind of ``resonance'', in a loose sense, when the length scales are commensurate and the layers can fit perfectly into the substrate patterns. Since the cholesteric pitch can be easily controlled with temperature or doping this might present interesting ways of controlling the optical properties or as a medium for colloidal templating \cite{Lintuvuori2013}.

%
%

\end{document}